\documentclass{article}
\usepackage{ijcai25}

\usepackage{times}
\usepackage{soul}
\usepackage{url}
\usepackage[hidelinks]{hyperref}
\usepackage[utf8]{inputenc}
\usepackage[small]{caption}
\usepackage{graphicx}
\usepackage{amsmath}
\usepackage{amsthm}
\usepackage{booktabs}
\usepackage{algorithm}
\usepackage{algorithmic}
\usepackage[switch]{lineno}
\usepackage{color}

\usepackage{natbib}
\usepackage{amssymb}
\usepackage{multirow}

\title{Model-Based Closed-Loop Control Algorithm for Stochastic Partial Differential Equation Control}

\author{
    Author Name
    \affiliations
    Affiliation
    \emails
    email@example.com
}

\author{
Peiyan Hu$^1$
\and
Haodong Feng$^3$\and
Yue Wang$^2$\footnote{Corresponding author.\\
Published as a conference paper of International Joint Conferences on Artificial Intelligence 2025.}\And
Zhiming Ma$^1$\\
\affiliations
$^1$Academy of Mathematics and Systems Science, Chinese Academy of Sciences\\
$^2$Zhongguancun Academy\\
$^3$Westlake University
\emails
hupeiyan18@mails.ucas.ac.cn,
fenghaodong@westlake.edu.cn,
yuewang@bjzgca.edu.cn,
mazm@amt.ac.cn
}

\begin{document}

\maketitle

\begin{abstract}
Neural operators have demonstrated promise in modeling and controlling systems governed by Partial Differential Equations (PDEs).
Beyond PDEs, Stochastic Partial Differential Equations (SPDEs) play a critical role in modeling systems influenced by randomness, with applications in finance, physics, and beyond.  However, controlling SPDE-governed systems remains a significant challenge. On the one hand, the regularity of the system's state (which can be intuitively understood as smoothness) deteriorates, making modeling and generalization more challenging. On the other hand, this stochasticity also renders control more unstable and thus less accurate.
To address this gap, we propose the Model-Based Closed-Loop Control Algorithm (MB-CC), the first model-based closed-loop control method for SPDEs. 
MB-CC introduces two key innovations to enhance control robustness and efficiency: a Regularity Feature (RF) block and a closed-loop strategy with an operator-encoded policy network. The RF block, inspired by the regularity structure theory of SPDEs, addresses noise-induced irregularities by transforming the network's input—including the system state and noise-perturbed external forces—into a refined feature space for improved forward prediction.  Compared to previous works using regularity features, we introduce a new parameterization, data augmentation, and extend the RF block as a plug-and-play component. Additionally, to achieve closed-loop control, we introduce an operator-encoded policy network to map the current state to optimal control, which integrates physical priors and swiftly makes decisions based on states returned by the environment.
We conduct a systematic evaluation of MB-CC on two notable SPDEs, showcasing its effectiveness and efficiency. The ablation studies show its ability to handle stochasticity more effectively.

\end{abstract}

\section{Introduction}
The simulation and control of stochastic partial differential equations (SPDEs) are essential in both scientific and engineering areas due to SPDEs' ability to model dynamic systems with stochasticity. For instance, SPDEs model the evolution of financial derivatives and asset prices, incorporating market volatility and uncertainty, in quantitative finance \citep{baudoin2002conditioned, mccauley2013stochastic, braumann2019introduction}. The stochastic Navier-Stokes equation is widely used to describe turbulence, including ocean currents and atmosphere \citep{m04, d02, s96}. These applications demonstrate the importance of SPDEs in accurately representing dynamic systems influenced by stochastic processes \citep{t19, c17, h01, r82}. 

Meanwhile, with rapid developments of deep learning, solving and controlling PDEs and SPDEs with neural networks is more and more popular. On the one hand, the combination of deep learning and control raises both the speed and accuracy \citep{h22, p20, q22} compared with traditional methods. On the other hand, models for SPDEs are constructed specifically under the consideration of normal models' failure to handle the coarse noise \citep{c21, hu2022neural, g23}. 

Although deep-learning-based methods are proposed for PDE control, most existing methods are open-loop and thus lack accuracy \citep{y21, p21, h22, y22}. Also, despite all these emerging methods, it is difficult to apply these approaches directly to SPDEs' control. As mentioned before, random forcing is a common situation while solving control problems, however, the low regularity of random forcing, affecting both the control's stability and learning's accuracy, renders it a challenging issue, which leads to the necessity of developing control methods aimed at SPDEs. 

In this work, we propose a Model-Based Closed-Loop Control Algorithm (MB-CC) specifically for efficient and robust SPDE control. Given the increasing demand for the accuracy, robustness, and stability of control algorithms, we consider this problem from two perspectives: one is how to better model the forward dynamics, and the other is how to achieve greater stability and accuracy in control. Firstly, we design a modified RF block in order to develop a more advanced and robust neural operator for the following control task, overcoming the stochastic challenge. This block enables the model to better handle the low regularity caused by noise terms, and also embeds physical information more effectively, thereby enhancing generalization ability.
{Secondly, in order to enhance control efficiency and robustness, we then propose an operator-encoded policy net using the differentiability of the RF block. The introduction of an operator-encoded policy net incorporates the physical information contained in the operator and offers the benefit of feedback, enabling real-time adjustments for improved accuracy and stability in response to dynamic changes. These aspects result in the accurate and fast Model-Based Closed-Loop Control Algorithm (MB-CC). Furthermore, this framework can be readily integrated with commonly utilized network architectures, demonstrating its versatility and broad applicability. As for experiments, we choose the famous stochastic 1-D reaction-diffusion equation and 2-D Navier-Stokes equation, and evaluate MB-CC and baselines on tracking problems, which impose higher demands on the generalization ability. Our contributions can be summarized as follows:

\begin{itemize}
    \item \textbf{Modeling inspired by stochasticity}: In order to more accurately model the nonlinear dynamics affected by stochasticity, we introduce the Regularity Feature (RF) block, which is able to deal with the low regularity of stochastic noise in SPDEs specifically. Furthermore, we design the data augmentation and enhance its compatibility with various backbones. Additionally, its differentiability is essential in the training of the policy net.
    \item \textbf{Closed-loop control}: Different from the open-loop methods before, the operator-encoded policy net is adopted in our method. Firstly, it helps the algorithm better handle stochasticity because it encodes the physical information and can adjust the next control action based on the current state affected by the noise. Secondly, it eliminates the need for optimization during control, allowing for the generation of control sequences in a very short time.
    \item \textbf{Experiments:} We conduct experiments on important and common equations, including the 1-D stochastic reaction-diffusion equation and the 2-D stochastic Navier-Stokes equation. The results demonstrate that MB-CC can significantly improve control outcomes, including both accuracy and speed.
\end{itemize}

\section{Related work}

\paragraph{AI for (S)PDE simulation} There are two categories of methods that utilize deep learning techniques to simulate PDEs. Some works take neural networks to directly approximate the solution function by training them with physical losses, such as the residual of PDEs or the modified variational form of residual, that represent the PDEs \citep{r19, y18}. Others focus on designing neural operators that can learn operators from the problem functions to the solution functions to solve a series of parametric PDEs \citep{li20, l21, li_gno_20, t22}. For SPDEs, taking into account the unique characteristics of inherent noise, several studies have proposed neural operators specifically tailored to solve SPDEs. Neural SPDEs propose a model to simulate SPDEs with the consideration of stochasticity \citep{n22}, while the DLR-Net utilizes regularity features to enhance the performance \citep{hu2022neural,g23}.

\paragraph{Traditional PDEs control methods} In the control theory, the functional gradient is obtained through the Fréchet derivatives whose calculation is quite computationally expensive. Based on the previous work \citep{lions12}, \citet{c74} proposes the adjoint method based on the adjoint state to compute the functional gradient without the need for Fréchet derivatives, which is a mainstream way for PDE control problems. Other recent works improve the efficiency of Fréchet derivative's calculation \citep{zhou2024solving}. Furthermore, another well-known control method is Proportional Integral Derivative (PID) \citep{ang2005pid, 1580152} control, which controls a system by continuously calculating and adjusting the control input to minimize the difference between a desired state and the actual process. As a common single-input single-output (SISO) control method, the PID is difficult to directly apply to multiple-input multiple-output (MIMO) systems. Its application to the MIMO system often requires additional decoupling and target planning modules. Therefore, although effective in some scenarios, it is only applicable to a limited range of problems.

\paragraph{AI for (S)PDE control} Compared with traditional methods, deep learning dramatically speeds up the process of control. Most works focus on PDE-constrained control problems.
Some researchers propose a hierarchical scheme consisting of a predictor network and a control network \citep{p20}. Another approach consists of two stages: a surrogate model is trained in the first stage and the control is set as a learnable parameter in the second stage \citep{h22}. Besides, \citet{wei2024generative} and \citet{hu2024wavelet} propose a generative control method that generates the control sequences and state trajectories simultaneously. There is also an open-source project with various control environments and learning-based controllers \citep{zhang2023controlgym}. However, these approaches are not directly applicable to SPDEs, as the stochastic nature introduces high complexity and difficulty in addressing control issues. 
For systems with inherent stochasticity, there has been little work focusing on their control problems. \citet{zhang2022neural} address the control of SDEs by introducing the exponential stabilizer (ES) based on stochastic Lyapunov theory and the asymptotic stabilizer (AS) based on stochastic asymptotic stability theory. \citet{pirmorad2021deep} considers SPDEs but directly uses the deep deterministic policy gradient (DDPG) algorithm from reinforcement learning, only making a preliminary attempt with the stochastic Burgers’ equation. In contrast, our work specifically designs the algorithm to address the stochasticity of SPDE systems and conducts more comprehensive experiments.

\section{Preliminary}

In this section, we introduce the background and notations used throughout this work.

\subsection{Regularity Structure of SPDEs}
In this work, we consider an SPDE on $[0,T]\times D$ as follows
\begin{align}
    \partial_tu-\mathcal{L}u = \mu(u,\partial_1u,\cdots &,\partial_du)+f+\sigma(u,\partial_1u,\cdots,\partial_du)\xi,\nonumber\\
    & u(0,x)=u_0(x), \label{SPDE}
\end{align}
where $x\in D\subset\mathbb{R}^d$, $t\in[0,T]$, $\mathcal{L}$ is a linear differential operator, $f$ is the deterministic forcing term, $\xi$ is the random forcing, $u_0:D\rightarrow\mathbb{R}$ is the initial condition. $u$, $f$ is in the Banach space $\mathcal{U}$ and $\mathcal{F}$ respectively.

Under local Lipschitz condition on $\mu,\sigma$ with respect to suitable norm, this SPDE has a unique mild solution \citep{h14, c21}:
\begin{align} 
    u_t=e^{t\mathcal{L}}u_0+\int_0^t e^{(t-s)\mathcal{L}}\mu(u_s,\partial_1u_s,\cdots ,\partial_du_s) ds \nonumber\\
    +\int_0^t  e^{(t-s)\mathcal{L}}(f+\sigma(u_s,\partial_1u_s,\cdots,\partial_du_s)\xi) ds. \label{mild}
\end{align}

According to the representation of the mild solution above, we define two linear operators $I[f](t)=\int_{0}^te^{(t-s)\mathcal{L}} f(s)ds$ and $I_c[u_0](t)=e^{t\mathcal{L}}u_0$ 
for any function defined on $[0,T]\times D$ to $\mathbb{R}^d$. 
Like in PDEs,  using Picard theorem, we can get the following recursive sequence that can approximate the solution $u$ of equation (\ref{SPDE}) as $n\rightarrow\infty$
{\small\begin{equation}
    \begin{aligned}
    &u^0_t = I_c[u_0]_t,\quad
    u^{n+1}_t = I_c[u_0]_t+I[\mu(u^n)+f+\sigma(u^n)\xi]_t. 
    \end{aligned} \label{iteration}
\end{equation}}
Then using Taylor expansion, another recursive sequence approximates $u$ as $m,l,n\rightarrow\infty$, where $\mu^{(k)}$ denotes the k-th derivative of $\mu$:
\begin{align}
    &u^{0,m,l}_t = I_c[u_0]_t,\\
\label{feature}
    &u^{n+1,m,l}_t = I_c[u_0]_t + \sum^m_{k=0}\frac{\mu^{(k)}(0)}{k!}I[(u^{n,m,l})^k]_t + I[f]_t \\ \notag
    &\quad\quad\quad + \sum^l_{k=0}\frac{\sigma^{(k)}(0)}{k!}I[(u^{n,m,l})^k\xi]_t. 
\end{align}
It is then revealed that the solution of SPDE can be approximated by a weighted sum of the features $I[f]$, $I[(u^{n,m,l})^k], I[(u^{n,m,l})^k\xi], l=0,\cdots,k; m=0,\cdots,k$.

\subsection{Problem Setup}

In this work, we aim to solve the control problem optimally and efficiently. Considering the stochastic dynamical system in the equation (\ref{SPDE}), we can define the tracking error of the SPDE-constraint problem as follows: 
\begin{align}
    &e(u_0,u^*,f,\xi) \notag \\
    =& e_{track}(u_0,u^*,f,\xi)+e_{energy}(f) \notag \\ 
    =&||u_t(u_0, f, \xi)-u^*||_{L^2((0,T]\times D)} + \alpha||f||_{L^2([0,T]\times D)}, \notag
\end{align}
where $u_0$ is the initial condition, $u^*$ is the function describing the target state of the system, $f$ is the deterministic external forcing that we can control and apply to the system, $\xi$ is the random noise (usually space-time white noise), $u_t$ is the time-dependent states depending on $u_0, f$ and $\xi$, and $\alpha$ is the coefficient used to adjust the weight of two terms. The first term constrains the system to the target state, while the second limits the norm of external forcing. Intuitively, the problem is to control the system state approach and track the target using minimal external force. Notably, this problem is challenging because it requires the system to reach the target state as quickly as possible, a scenario that rarely appears in the training set. 

Due to the appearance of random forcing, we consider the error in the sense of expectation:
\begin{align*}
    \hat{e}(u_0, u^*, f) = \underset{\xi}{\mathbb{E}}[e(u_0, u^*, f, \xi)].
\end{align*}
Thus the tracking problem is formally described as
\begin{align}
    \underset{f\in \mathcal{F}}{\rm min} \it\ \hat{e}(u_0,u^*,f).
\end{align}

\section{Methodology}

\begin{figure}
  \centering
  \includegraphics[width=0.95\columnwidth]{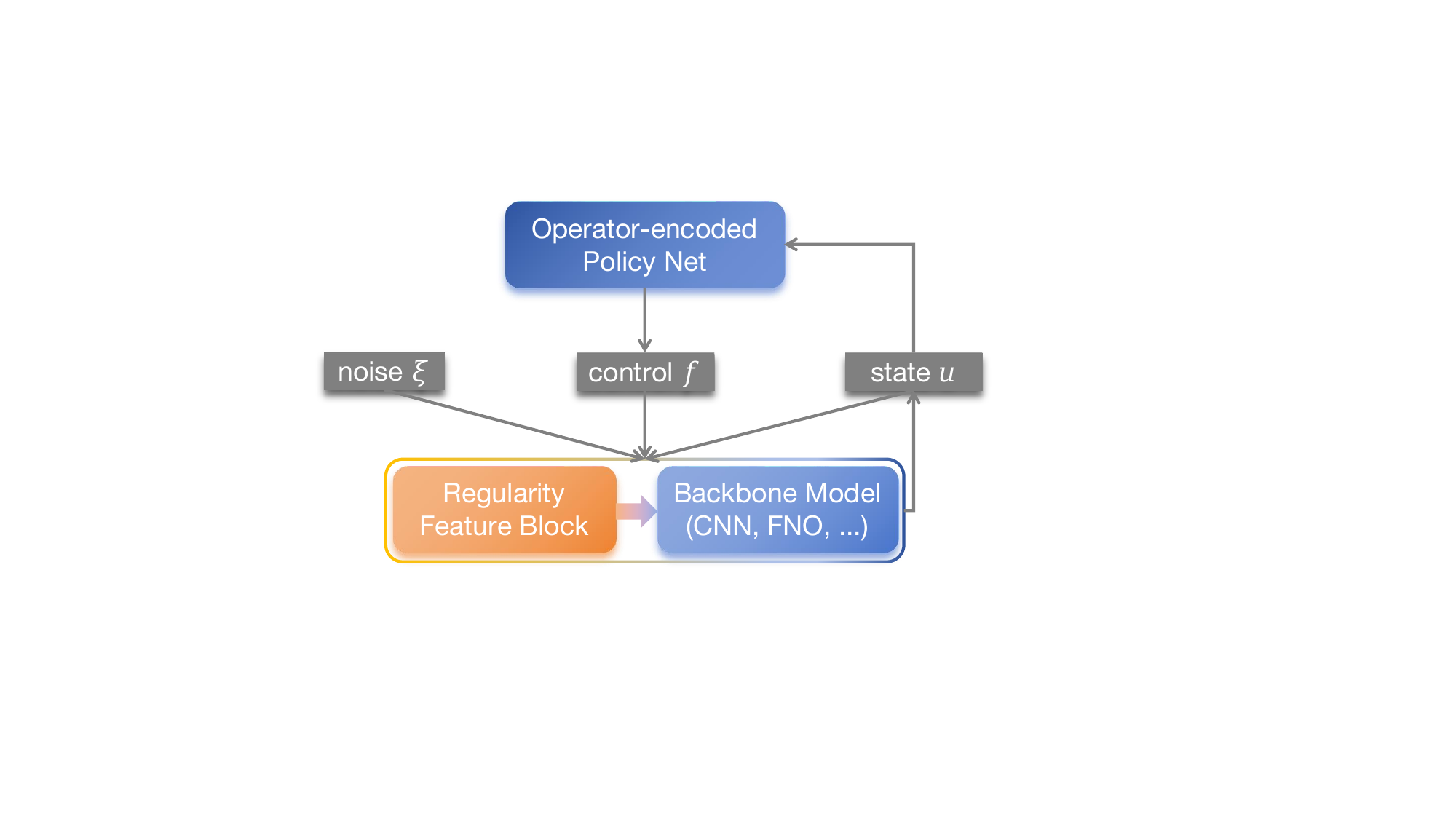}
  \caption{\textbf{Overview of the Model-Based Closed-Loop Control Algorithm (MB-CC).}}
  \label{fig:mb-cc}
\end{figure}

In this section, we will introduce our proposed Model-Based Closed-Loop Control Algorithm (MB-CC) as shown in Figure \ref{fig:mb-cc}, which has the following two parts: \emph{modeling with our design RF block} and \emph{operator-encoded control}.

\subsection{Modeling Inspired by Stochasticity}

\begin{figure}[b]
  \centering
  \includegraphics[width=1\columnwidth]{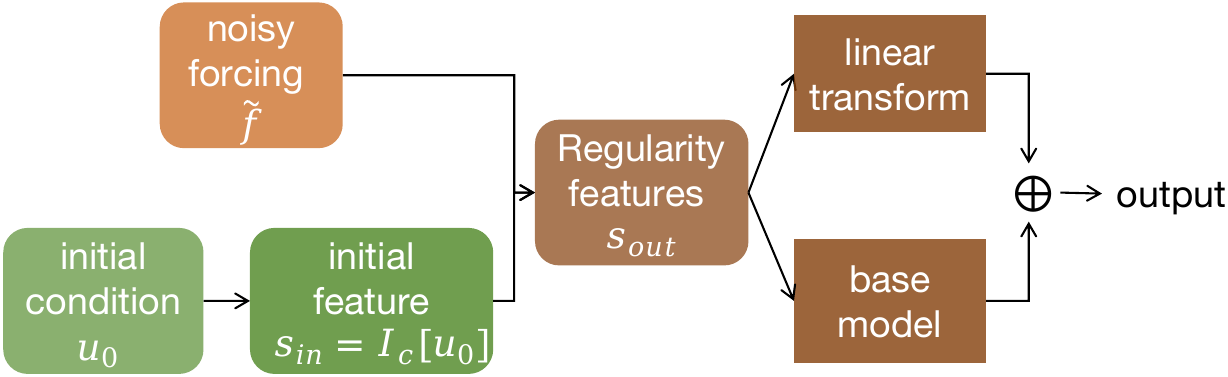}
  \caption{\textbf{Overall architecture of the combination of the RF block and the base model.}}
  \label{fig:rfblock}
\end{figure}

\begin{figure*}
  \centering
  \makebox[\textwidth][c]{\includegraphics[width=2.2\columnwidth]{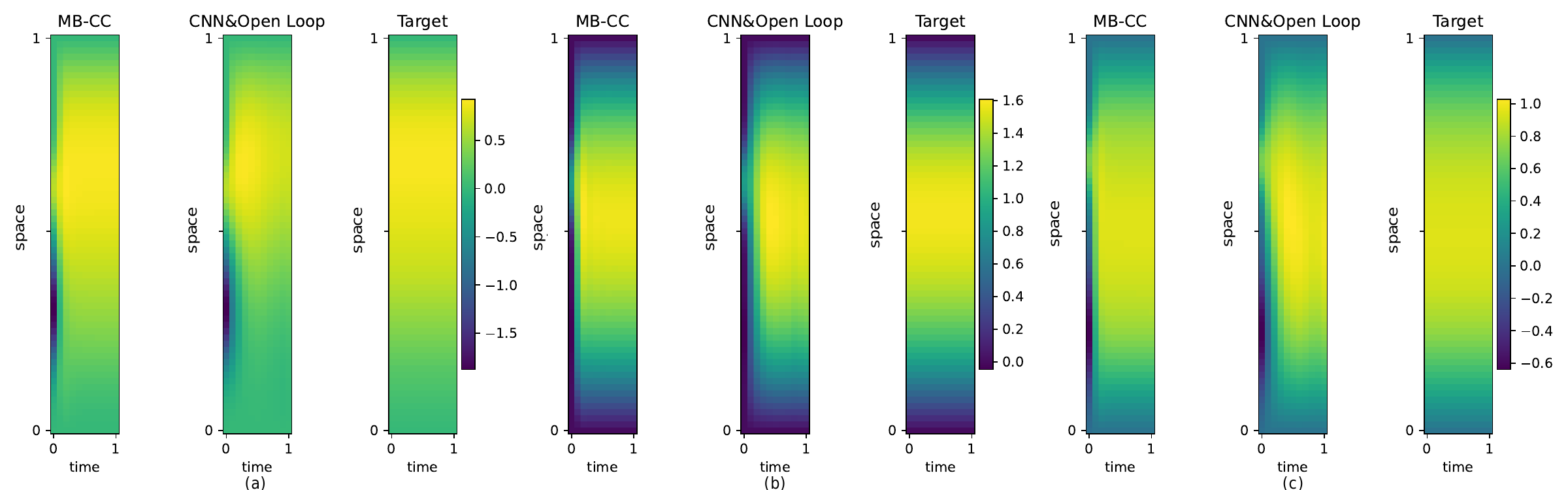}}
  \caption{\textbf{Visualization of results on the stochastic reaction-diffusion equation.} The figure shows the visualized results of controlling three samples using MB-CC, results using CNN \& OpenLoop, and the control targets. It is obvious that the results controlled by MB-CC are significantly closer to the target.}
  \label{fig:rd}
\end{figure*}

The architecture of the entire model, which learns the forward dynamics, consists of two components. 
\paragraph{Regularity Structure Features.} Due to the prominence of the regularity structure theory in the analysis of SPDEs \citep{h14}, we choose regularity structure vectors as features for the model. We incorporate the RF block, which maps the initial condition $u_0$ and forcing $\Tilde{f}$ (the collection of the deterministic forcing $f$ and random forcing $\xi$) to regularity structure vectors. 

These regularity structure vectors form the feature set $\mathcal{S}_{n,m,l}$, and are generated and computed iteratively as Algorithm \ref{alg:mfgen}.
The outer loop corresponds to the number of Picard iterations. In each Picard iteration, we first compute the set of integrands, $\mathcal{Z}$, that appear in the Taylor expansion, followed by a discrete time-step iteration to calculate the time integral of each element in $\mathcal{Z}$.

\begin{algorithm}[ht]
\caption{Generation of Regularity Structure Features}
\label{alg:mfgen} 
\textbf{Input}: Initial feature $s^{in}$, forcing $\Tilde{f}$\\
\textbf{Parameter}: Height $n,m,l$, discretized operator $\mathcal{L}^{\text{dis}}$, time grid points $t_0,\cdots,t_{K-1}$\\
\textbf{Output}: $s^{out}=\mathcal{S}_{n,m,l}(s^{in},\Tilde{f})$
\begin{algorithmic}[1]
\STATE Initial function set $\mathcal{S}_{0,m,l}=\{s^{in}\}$
\FOR {$p=1,\cdots,n$}
\STATE  Generate set $\mathcal{Z}_{p,m,l}=\{\Tilde{f}^j\prod_{i=1}^k\partial^{a_i}s_i: s_i\in \mathcal{S}_{p-1,m,l},a_i, j\in\{0,1\}, k \in \mathbb{N}, 1\leq k+j\leq mI_{j=0}+\ell I_{j>0}\}$
\FOR {$k=0,1,\cdots,K-1$}
\STATE For $z\in \mathcal{Z}_{p,m,l}$, $I[z]_{t_{k+1}}=(I[z]_{t_k}+z_{t_k}\cdot\delta _t)\cdot (Id-\mathcal{L}^{\text{dis}}\cdot\delta t)^{-1}$
\ENDFOR
\STATE $\mathcal{S}_{p,m,l}=\{I[z],z\in\mathcal{Z}_{p,m,l}\}\cup\mathcal{S}_{p-1,m,l}$ 
\ENDFOR
\end{algorithmic}
\end{algorithm} 

The parameters $n, m$, and $l$ are hyperparameters we select, which affect the number of vectors in $\mathcal{S}_{n,m,l}$. The initial feature $s^{\text{in}}=I_c[u_0]$, the final output is $s^{\text{out}}= \mathcal{S}_{n,m,l}(s^{\text{in}},\Tilde{f})$, which contain $N_{\mathcal{S}}$ features ($N_\mathcal{S}$ is decided by the hyperparameters $n,m$ and $l$). The discretized operator $\mathcal{L}^{\text{dis}}$ is the discretization of the operator in the SPDE from the equation (\ref{SPDE}), implemented using the finite difference numerical method. We provide an example of $\mathcal{L}^{\text{dis}}$ in Appendix D.

These features enable the model to better handle the issue of low regularity caused by noise terms, and it embeds the crucial physical information of the operator, thereby enhancing the model’s ability to represent stochastic systems. 

\paragraph{Model Architecture.} We base our design on equation (\ref{feature}) to reasonably and effectively integrate the features with the backbone network. According to the equation (\ref{feature}), the linear combination of the regularity structure features is actually the first $N_{\mathcal{S}}$ term of the Taylor expansion. In consequence, we let the backbone neural network $W_{\theta}$ with weights $\theta$ only be used to approximate the residual part caused by the truncation of the Taylor expansion, which means the output $\Tilde{u}$ of the entire model can be represented as
\begin{align}
    \Tilde{u}_\theta=\theta_1 s^{\text{out}}+ W_{\theta_2}(s^{\text{out}},\mathcal{O}),
\end{align}
where $\mathcal{O}$ is the concatenation of the space and time grid, $\theta_1\in \mathbb{R}^{N_{\mathcal{S}}}$ and $\theta_2$ are the weights learned through training.

\paragraph{Data Augmentation.} In addition, the transformation from the initial condition and forcing to regularity features causes the data of different time steps to separate from each other, which makes the model with the RF block have difficulty fitting data far away from the most. As a result, according to the training error of the model trained with raw data, we address this issue by increasing the proportion of difficult-to-fit data in the dataset through duplication.

\paragraph{Plug-and-Play Component.} Besides, we extend the RF block as a plug-and-play component, which is part of our proposed algorithm framework. The backbone model can be changed to any suitable neural network architecture, for instance, using either Convolutional Neural Network (CNN) \citep{o2015introduction}, Fourier Neural Operator (FNO) \citep{li20}, or other neural operators. In our experiments, we test this component in combination with different network architectures. Results show that, regardless of the backbone network model used, our proposed framework achieves significant improvements.

\subsection{Closed-Loop Operator-Encoded Control}
\paragraph{Closed-Loop Control.} Open-loop control algorithms function without utilizing feedback to adjust their inputs, resulting in a significant drawback: the absence of error correction. This limitation leads to reduced accuracy and adaptability in response to changes or disturbances in the system, particularly in stochastic systems. 
To overcome the aforementioned limitations of open-loop control, we present a policy net, inspired by reinforcement learning algorithms \citep{lillicrap2015continuous, schulman2017proximal, haarnoja2018soft}, that maps from the state to the optimal control in order to decide the next control action based on the current state in the closed-loop control manner.

\paragraph{Operator-Encoded Policy Net $P_{\gamma}$.} By rearranging the terms in Equation 1, we can obtain
\begin{align}
    f = \partial_tu-\mathcal{L}u &- \mu(u,\partial_1u,\cdots,\partial_du) \notag \\
    &- \sigma(u,\partial_1u,\cdots,\partial_du)\xi. \nonumber
\end{align}
From this, it can be seen that the external force term \( f \) involves \( \mathcal{L}u \), which inspires us to embed operator information into the design of the policy net. We use $\mathcal{L}^{\text{dis}}$ as introduced in the previous section. For the current state $u_t$, we concatenate $u_t$ and $\mathcal{L}^{\text{dis}}u_t$, creating a tensor embedded with operator information that is used as input of the policy net. For the target state $u_T$, we perform the same operation. Due to the inclusion of SPDE physical information in the operator, as introduced in the previous section, we not only enhance the network’s performance but also significantly improve its generalization ability by encoding key physical information.

\paragraph{Loss Function.} Fed into the current state $u_t$, target state $u_T$ and time $t$, the policy net $P_\gamma(u_t,u_T,t)$ outputs the action $f_t$. As the optimization problem is in the sense of expectation, we use the mean of multiple samples to approximate the expectation. Consequently, we generate enough number of different random noise $\xi_i(i=1,\cdots,N)$ and calculate the mean of results using the model's predicted states $u_\theta$. %
The loss function of the policy net is then defined as
\begin{align}
    &L(u_0, u_T) \notag\\
    &= \frac{1}{N}\sum^N_{i=1}(||\Tilde{u}_{t, \theta}(u_0, P_\gamma(\Tilde{u}_t,u_T,t), \xi_i)\nonumber-u_T||_{L^2((0,T]\times D)}\\
    &+ \alpha||P_\gamma(\Tilde{u}_t,u_T,t)||_{L^2([0,T]\times D)} ), \label{track_loss}
\end{align} 
where $\gamma$ is the learnable weights of the policy net, $\Tilde{u}_{t, \theta}(u_0, P_\gamma(\Tilde{u}_t,u_T,t), \xi_i)$ is simulation of the physical system's states. %

\paragraph{Data.} Our work is data-efficient, demonstrating in two aspects. Firstly, the training data of the policy net $P_\gamma$ are same as that used to train the operator model before, which means no additional data are needed in training the policy net. Secondly, we can even only use $u_0$ and $u_T$ random sampled from the data distribution without the force term $f$. Therefore, throughout the process, we use the model's prediction $\Tilde{u}_\theta$ to simulate the trajectories rather than interact with the environment, as it is supposed to be close enough to the environment.

\section{Experiment}

\begin{table}
\centering
\begin{tabular}{l|c}
\toprule
Model & Error \\
\midrule
RF-CNN            & 0.0115 = 0.0106 + 0.0003 + 0.0002 +
\underline{\textbf{0.0003}}\\
CNN               & 0.0182 = 0.0163 + 0.0010 + 0.0004 + \underline{0.0006}\\
RF-FNO            & \textbf{0.0029} = 0.0001 + 0.0004 + 0.0003 +
\underline{0.0021}\\
FNO               & 0.0138 = 0.0001 + 0.0020 + 0.0014 +
 \underline{0.0104}\\
\bottomrule
\end{tabular}
\caption{\textbf{Results of modeling the stochastic reaction-diffusion equation.} The four terms of the relative error are reconstruction error of $f, u_0, u_1$, and prediction error (underline). The best results are highlighted in bold.}
\label{tab:rdmodel}
\end{table}

In this section, we aim to address the following question: Do the two major parts of MB-CC contribute to improve the performance of SPDEs control? Specifically, can they effectively improve the handling of noise? Therefore, we evaluate our method on the tracking problem of two widely-used SPDEs, including the stochastic reaction-diffusion equation with linear multiplicative forcing and the 2-D stochastic Navier-Stokes equation with additive noise. On each equation, fifty tasks are tested to provide sufficient evidence of the answer. The objective of each task is sampled from the distribution of the system state at the last time step in the training data. The tracking problem aims for the algorithm to output control signals that keep the system’s state close to the target at all times, which is a challenging out-of-distribution problem. We provide our code in the supplementary materials.

The performance of modeling and control is evaluated and reported, respectively. Through the experiments, we respectively take CNN \citep{o2015introduction} and FNO \citep{li20} as the backbone of our proposed method with RF block. The naive CNN and FNO are applied as the baseline methods to show the improvement using RF block while demonstrating its plug-and-play ability. Specifically, following the methodology of the previous work \citep{h22}, we employ these two models as auto-regressive models to learn the state transitions between single steps. Additionally, the models simultaneously output the input \(f\) and \(u_0\) to calculate the reconstruction loss. To show the comprehensive advantages of our framework, we apply open-loop control using CNN, FNO, RF-CNN, RF-FNO, and Soft Actor-Critic (SAC) \citep{haarnoja2018soft} as baseline methods to show the superior performance of our proposed policy net with respect to three metrics: $L_2$ relative error, objective loss $\hat{e}$, and time.

\subsection{1-D Stochastic Reaction-Diffusion Equation}

\begin{table}
\centering
\begin{tabular}{l|c|c|c|c}
\toprule
Model & Method & $\hat{e}$ & $\hat{e}_{track}$ & $\hat{e}_{energy}$ \\
\midrule
RF-CNN            & Open Loop & 0.0780 & 0.0504 & 0.0276 \\
\textbf{RF-CNN}   & \textbf{Policy Net} & \textbf{0.0658} & \textbf{0.0359} & 0.0299 \\
CNN               & Open Loop & 0.2035 & 0.1872 & 0.0163 \\
CNN               & Policy Net & 0.1258  & 0.1003 & 0.0255 \\
RF-FNO            & Open Loop & 0.1284 & 0.1002 & 0.0282 \\
\textbf{RF-FNO}   & \textbf{Policy Net} & 0.1126 & 0.0866 & 0.0260 \\
FNO               & Open Loop & 0.5865 & 0.5711 & 0.0153 \\
FNO               & Policy Net & 0.1208 & 0.0879 & 0.0329 \\
SAC               & Policy Net & 2.5924 & 2.5631 & 0.0292 \\
\bottomrule
\end{tabular}
\caption{\textbf{Results of tracking problems on the stochastic reaction-diffusion equation.} The table records the objective loss $\hat{e}$, consisting of the tracking loss $\hat{e}_{track}$, which measures the distance of the trajectory and the target state, and the energy loss $\hat{e}_{energy}$. The best results are highlighted in bold.}
\label{tab:rdtrack}
\end{table}

\begin{table}
\centering
\begin{tabular}{l|c|c}
\toprule
Model & Open Loop & Policy Net\\
\midrule
RF-CNN       & 367.40 & 0.25 \\
CNN          & 54.35 & 0.31 \\
RF-FNO       & 67.83 & 0.30 \\
FNO          & 44.10 & 0.25 \\
\bottomrule
\end{tabular}
\caption{\textbf{Mean inference time on the stochastic reaction-diffusion equation.} The unit is seconds (s).}
\label{tab:rdtime}
\end{table}

We first evaluate our method on the reaction-diffusion equation with multiplicative forcing as in \citep{i21} with the Dirichlet boundary, which is critical in many fields, including environmental science, energy development, and fluid mechanics. The form is given by
{\begin{equation}
    \begin{aligned}
    \partial_t u-\nu\Delta u &= 3u-u^3+f+\sigma u \xi,\quad(t,x)\in[0,T]\times D\\
    u(t,x) &= 0,\quad(t,x)\in[0,T]\times\partial D\\
    u(0,x) &= u_0(x),
    \end{aligned} 
\end{equation}}
where $\nu=0.1$ is the viscosity parameter, $\xi$ is the smoothed space-time white noise scaled by $\sigma = 0.05$. We first generate the space-time white noise using the numerical simulator as the previous work \citep{i21}, the noise is then smoothed using the moving average algorithm with window size 3. We choose $D=[0,1]$, $T=1$, and take 64 space grid points and 11 time grid points uniformly. Besides, $u_0(x)$ and $f$ are randomly sampled from a distribution following the previous work \citep{h22}. 

As for these models' training, we train RF-CNN, CNN, RF-FNO, and FNO with 4000 trajectories, respectively, while testing on 500 trajectories. The relative $L_2$ errors of CNN, RF-CNN, FNO, and RF-FNO are shown in Table \ref{tab:rdmodel}, from which it is obvious that models with the RF block have much more accuracy of prediction, which verifies that our introduced modeling inspired by stochasticity can enhance the model's ability to handle the complex stochastic system with low regularity.

\begin{table}
\centering
\begin{tabular}{l|c}
\toprule
Model & Error \\
\midrule
RF-CNN            & 0.0326 = 0.0050 + 0.0122 + 0.0057 + \underline{0.0096}\\
CNN               & 0.0784 = 0.0154 + 0.0327 + 0.0068 + \underline{0.0235}\\
RF-FNO            & \textbf{0.0061} = 0.0004 + 0.0000 + 0.0000 + \underline{\textbf{0.0056}}\\
FNO               & 0.0124 = 0.0010 + 0.0006 + 0.0005 + \underline{0.0103}\\
\bottomrule
\end{tabular}
\caption{\textbf{Results of modeling the stochastic Navier-Stokes equation.} We train RF-CNN, CNN, RF-FNO, and FNO with 400 trajectories respectively, while testing on 500 trajectories. The four terms of the relative error are reconstruction error of $f, u_0, u_1$, and prediction error (underline). The best results are highlighted in bold.}
\label{tab:NSmodel}
\end{table}

As for control, we take 4000 trajectories to train the policy net. As mentioned before, the objective function of the tracking problem is defined as equation (\ref{track_loss}). In this setting, we take $\alpha=0.01$ and $N=50$. Figure \ref{fig:rd} presents the visualizations of the system controlled by MB-CC and open-loop CNN, from which we can observe that MB-CC's results are much closer to the target states. More visualizations can be found in the supplementary material. Besides, we provide detailed results in Table \ref{tab:rdtrack}, including $\hat{e}, \hat{e}_{track}, \hat{e}_{energy}$. It clearly indicates that the introduction of both the RF block and the policy net can notably enhance the performance, which is a test of our previous statement. In addition, MB-CC, due to its specialized design for SPDE systems, is able to outperform the reinforcement learning SAC algorithm.

We also provide the meantime of solving these tracking tasks in Table \ref{tab:rdtime}, from which we can observe that the policy net incredibly speeds up the process of control.

\subsection{2-D Stochastic Navier-Stokes Equation}

Next, we consider the important 2-D Navier-Stokes equation for a viscous, incompressible fluid in vorticity form:
\begin{eqnarray}
    \partial_t w-\nu\Delta w &=& -u\cdot\nabla w+f+\sigma\xi,\,\, (t,x)\in[0,T]\times D 
    \label{equa:NS}\nonumber\\
    w(t, 0) &=& w(t, 1),\quad\text{(Periodic BC)} \\
    \omega(0,x)&=&\omega_0(x)\nonumber
\end{eqnarray}
where $u$ is the 2-D velocity field, $\omega=\nabla\times u$ is the vorticity, the viscosity parameter $\nu=0.02$, $\sigma=10^{-5}$, $T=1$, $D=[0,1]^2$. $\omega_0$ and $f$ are sampled from the 2-D form of the initial condition's and forcing's distribution in the reaction-diffusion equation. The smoothing algorithm of noise $\xi$ is the same as above. The generation of 2-D space-time white noise and the numerical solver follow \citep{c21}. When collecting the data, we take the pseudo-spectral method on $40 \times 40$ space grid and 200 time grid to solve the equation and then downsample the solution every 20 time steps.

\begin{table}
\centering
\begin{tabular}{l|c|c|c|c}
\toprule
Model & Method & $\hat{e}$ & $\hat{e}_{track}$ & $\hat{e}_{energy}$ \\
\midrule
RF-CNN            & Open Loop & 14.7297 & 13.9981 & 0.7316 \\
\textbf{RF-CNN}& \textbf{Policy Net} & 8.2239 & 7.6071 & 0.6168 \\
CNN               & Open Loop & 20.8153 & 20.3278 & 0.4875 \\
CNN               & Policy Net & 12.5431  & 12.0509 & 0.4921 \\
RF-FNO            & Open Loop & 7.8103 & 6.9416 & 0.8686 \\
\textbf{RF-FNO}& \textbf{Policy Net} & \textbf{1.7775} & \textbf{0.9253} & 0.8522 \\
FNO               & Open Loop & 44.8132 & 43.6540 & 1.1592 \\
FNO               & Policy Net & 2.9588 & 2.4274 & 0.5314 \\
SAC               & - & 121.2406 & 115.2638 & 5.9768 \\
\bottomrule
\end{tabular}
\caption{\textbf{Results of tracking problems on the stochastic Navier-Stokes equation.} The table records the objective loss $\hat{e}$, consisting of the tracking loss $\hat{e}_{track}$ and the energy loss $\hat{e}_{energy}$. The best results are highlighted in bold.}
\label{tab:NStrack}
\end{table}

\begin{table}
\centering
\begin{tabular}{l|c|c}
\toprule
Model & Open Loop & Policy Net\\
\midrule
RF-CNN       & 81.78 & 0.09 \\
CNN          & 59.44 & 0.02 \\
RF-FNO       & 29.64 & 0.10 \\
FNO          & 39.69 & 0.09 \\
\bottomrule
\end{tabular}
\caption{\textbf{Mean time of solving tracking problems on the stochastic Navier-Stokes equation.} The unit is seconds (s).}
\label{tab:NStime}
\end{table}

\begin{table*}
\centering
\begin{tabular}{l|c|c|c|c|c|c|c}
\hline
\multirow{2}{*}{Model} & \multirow{2}{*}{Method} & \multicolumn{3}{c|}{$\sigma=0.05$} & \multicolumn{3}{c}{$\sigma=1$} \\
\cline{3-8}
&& $\hat{e}$ & $\hat{e}_{track}$ & $\hat{e}_{energy}$ & $\hat{e}$ & $\hat{e}_{track}$ & $\hat{e}_{energy}$ \\
\midrule
RF-CNN            & Open Loop & 0.0782 & 0.0507 & 0.0276 & 0.5501 & 0.5150 & 0.0351 \\
\textbf{RF-CNN}   & \textbf{Policy Net} & 0.0658 & 0.0360 & 0.0298 & 0.0767 & 0.0460 & 0.0307 \\
CNN               & Open Loop & 0.2035 & 0.1872 & 0.0163 & 0.6533 & 0.6295 & 0.0238 \\
CNN               & Policy Net & 0.1258  & 0.1003 & 0.0255 & 0.1378 & 0.1115 & 0.0263 \\
\bottomrule
\end{tabular}
\caption{\textbf{Control results with different scales of the space-time white noise.} We choose noise scaled by $\sigma=0.05$ and 1, and test four methods on these two systems.}
\label{tab:ablation}
\end{table*}

For the learning of CNN, FNO, RF-CNN, and RF-FNO, we take 400 data, a smaller amount of data, to train and 500 data to test. Results in Table \ref{tab:NSmodel} further verify that the model improved by the RF block better captures and models systems with inherent stochasticity.

On this SPDE, we consider $\alpha=100$ and $N=20$. In addition, we again train the policy net with 4000 data. The detailed control results are reported in Table \ref{tab:NStrack}, from which it is obvious that our proposed MB-CC framework still achieves the best performance, which confirms that our proposed specialized design for SPDE is both reasonable and effective.

As for time efficiency, it can be observed that the policy network, compared with open-loop control, can still significantly reduce the time required for inference.

\vspace{-5pt}
\begin{figure}
  \centering
  \includegraphics[width=0.85\columnwidth]{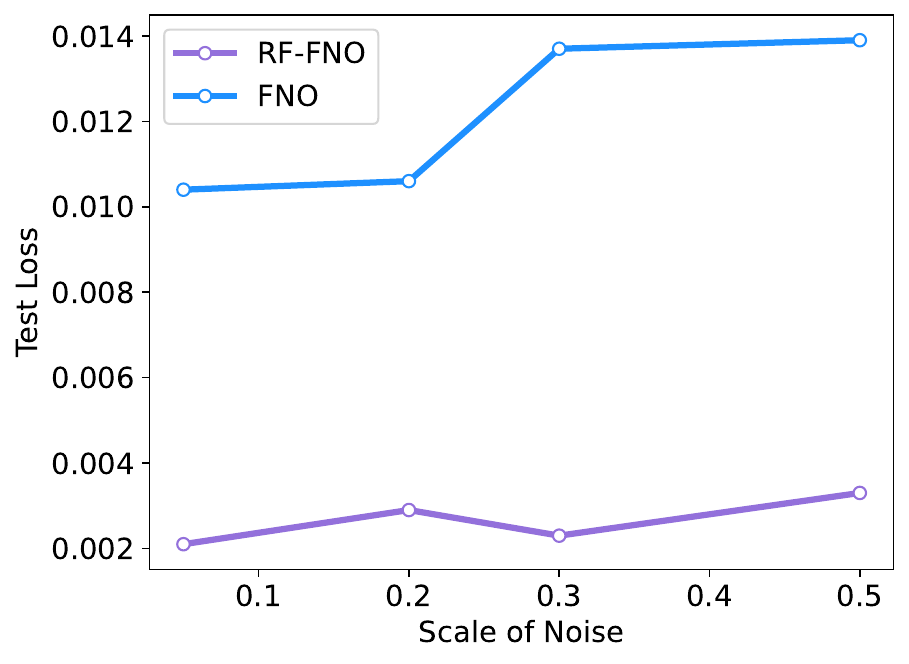}
  \vspace{-5pt}
  \caption{\textbf{Training and testing the forward models with different scales of the space-time white noise.}}
  \label{fig:noise}
\end{figure}

\subsection{Further Analysis}

In this subsection, we want to further analyze and discuss several questions: (1) Is the effectiveness of MB-CC related to the choice of the base model? (2) Does the improvement brought by MB-CC truly come from its ability to handle stochasticity more effectively? (3) Is the enhanced modeling capability caused by the RF block due to its improved handling of stochasticity?

To answer the first question, this essentially considers how well MB-CC generalizes across different base models. Referring to Table \ref{tab:rdtrack} and Table \ref{tab:NStrack}, we can see that MB-CC consistently shows significant performance improvements when CNN or FNO is used as the base model. Furthermore, if we control for variables and consider only the RF block or the operator-encoded policy net, both can independently enhance control performance.

As for the second question, we show the results of an ablation study on the 1-D stochastic reaction-diffusion equation to further validate our proposal that this framework is particularly suitable for systems with stochasticity. Since the stochasticity of the system is caused by the noise term $\xi$, we adjust the coefficient $\sigma$, which determines the scale of the system's noise. 

Therefore, we choose a higher $\sigma=1$ and directly test RF-CNN and CNN trained with the $\sigma=0.05$ dataset in both Open Loop and Policy Net scenarios again. This task requires the model to not only accurately simulate dynamics with higher noise levels but also make effective decisions and adjustments based on new states. Results in Table \ref{tab:ablation} demonstrate that compared to other methods, the approach combined with MB-CC shows more stable control results when $\sigma$ increases. This indicates that the introduction of MB-CC indeed enhances the algorithm's ability to handle stochasticity more effectively.

For the third question, we conduct another ablation study, where we choose higher $\sigma$ to train and test RF-FNO and FNO. Results with $\sigma=0.05, 0.2, 0.3, 0.5$ are plotted in Figure \ref{fig:noise}, showing that as $\sigma$ increases, the prediction error of FNO significantly rises, whereas RF-FNO consistently maintains a very low error level. This demonstrates the strong capability of the RF block to enhance the model's ability to handle stochasticity.

To sum up, both two ablation studies show that MB-CC indeed helps the algorithm better handle stochasticity, leading to significant improvements in SPDE control.

\section{Conclusion}

In this paper, we propose a novel control framework MB-CC to solve the SPDE-constraint control problems. The SPDE control is challenging, since the deterioration of the system's state regularity complicates modeling and generalization, while the stochasticity also makes open-loop control more unstable and less accurate. To address these challenges, we introduce two major components in MB-CC: stochasticity-inspired modeling based on regularity structure theory and closed-loop control achieved through the operator-encoded policy net. Compared with baselines, our method is evaluated on well-known SPDEs, including the 1-D stochastic reaction-diffusion equation and 2-D stochastic Navier-Stokes equation, and performs well in both forward modeling and control. In the future, we plan to combine MB-CC with more diverse and larger architectures to control more complex stochastic systems.

\bibliographystyle{named}
\bibliography{ijcai25}

\begin{thebibliography}{}

\bibitem[\protect\citeauthoryear{Ang \bgroup \em et al.\egroup }{2005}]{ang2005pid}
Kiam~Heong Ang, Gregory Chong, and Yun Li.
\newblock Pid control system analysis, design, and technology.
\newblock {\em IEEE transactions on control systems technology}, 13(4):559--576, 2005.

\bibitem[\protect\citeauthoryear{Baudoin}{2002}]{baudoin2002conditioned}
Fabrice Baudoin.
\newblock Conditioned stochastic differential equations: theory, examples and application to finance.
\newblock {\em Stochastic Processes and their Applications}, 100(1-2):109--145, 2002.

\bibitem[\protect\citeauthoryear{Braumann}{2019}]{braumann2019introduction}
Carlos~A Braumann.
\newblock {\em Introduction to stochastic differential equations with applications to modelling in biology and finance}.
\newblock John Wiley \& Sons, 2019.

\bibitem[\protect\citeauthoryear{Chavent}{1974}]{c74}
Guy Chavent.
\newblock Identification of functional parameters in partial differential equations.
\newblock In {\em Joint Automatic Control Conference}, number~12, pages 155--156, 1974.

\bibitem[\protect\citeauthoryear{Chevyrev \bgroup \em et al.\egroup }{2024}]{i21}
Ilya Chevyrev, Andris Gerasimovi{\v{c}}s, and Hendrik Weber.
\newblock Feature engineering with regularity structures.
\newblock {\em Journal of Scientific Computing}, 98(1):13, 2024.

\bibitem[\protect\citeauthoryear{Cristofol and Roques}{2017}]{c17}
Michel Cristofol and Lionel Roques.
\newblock Simultaneous determination of the drift and diffusion coefficients in stochastic differential equations.
\newblock {\em Inverse problems}, 33(9):095006, 2017.

\bibitem[\protect\citeauthoryear{Duan \bgroup \em et al.\egroup }{2002}]{d02}
Jinqiao Duan, Hongjung Gao, and Bj{\"o}rn Schmalfu{\ss}.
\newblock Stochastic dynamics of a coupled atmosphere--ocean model.
\newblock {\em Stochastics and Dynamics}, 2(03):357--380, 2002.

\bibitem[\protect\citeauthoryear{Gong \bgroup \em et al.\egroup }{2023}]{g23}
Shiqi Gong, Peiyan Hu, Qi~Meng, Yue Wang, Rongchan Zhu, Bingguang Chen, Zhiming Ma, Hao Ni, and Tie-Yan Liu.
\newblock Deep latent regularity network for modeling stochastic partial differential equations.
\newblock In {\em Proceedings of the AAAI Conference on Artificial Intelligence}, volume~37, pages 7740--7747, 2023.

\bibitem[\protect\citeauthoryear{Haarnoja \bgroup \em et al.\egroup }{2018}]{haarnoja2018soft}
Tuomas Haarnoja, Aurick Zhou, Pieter Abbeel, and Sergey Levine.
\newblock Soft actor-critic: Off-policy maximum entropy deep reinforcement learning with a stochastic actor.
\newblock In {\em International conference on machine learning}, pages 1861--1870. PMLR, 2018.

\bibitem[\protect\citeauthoryear{Hairer}{2014}]{h14}
M.~Hairer.
\newblock A theory of regularity structures.
\newblock {\em Invent. Math}, 198(2):269--504, 2014.

\bibitem[\protect\citeauthoryear{Higham}{2001}]{h01}
Desmond~J Higham.
\newblock An algorithmic introduction to numerical simulation of stochastic differential equations.
\newblock {\em SIAM review}, 43(3):525--546, 2001.

\bibitem[\protect\citeauthoryear{Holl \bgroup \em et al.\egroup }{2020}]{p20}
Philipp Holl, Vladlen Koltun, and Nils Thuerey.
\newblock Learning to control pdes with differentiable physics.
\newblock {\em CoRR}, abs/2001.07457, 2020.

\bibitem[\protect\citeauthoryear{Hu \bgroup \em et al.\egroup }{2022}]{hu2022neural}
Peiyan Hu, Qi~Meng, Bingguang Chen, Shiqi Gong, Yue Wang, Wei Chen, Rongchan Zhu, Zhi-Ming Ma, and Tie-Yan Liu.
\newblock Neural operator with regularity structure for modeling dynamics driven by spdes.
\newblock {\em arXiv preprint arXiv:2204.06255}, 2022.

\bibitem[\protect\citeauthoryear{Hu \bgroup \em et al.\egroup }{2024}]{hu2024wavelet}
Peiyan Hu, Rui Wang, Xiang Zheng, Tao Zhang, Haodong Feng, Ruiqi Feng, Long Wei, Yue Wang, Zhi-Ming Ma, and Tailin Wu.
\newblock Wavelet diffusion neural operator.
\newblock {\em The Thirteenth International Conference on Learning Representations}, 2024.

\bibitem[\protect\citeauthoryear{Hwang \bgroup \em et al.\egroup }{2021}]{h22}
Rakhoon Hwang, Jae~Yong Lee, Jin~Young Shin, and Hyung~Ju Hwang.
\newblock Solving pde-constrained control problems using operator learning, 2021.

\bibitem[\protect\citeauthoryear{Li \bgroup \em et al.\egroup }{2006}]{1580152}
Yun Li, Kiam~Heong Ang, and G.C.Y. Chong.
\newblock Pid control system analysis and design.
\newblock {\em IEEE Control Systems Magazine}, 26(1):32--41, 2006.

\bibitem[\protect\citeauthoryear{Li \bgroup \em et al.\egroup }{2020}]{li_gno_20}
Zongyi Li, Nikola Kovachki, Kamyar Azizzadenesheli, Burigede Liu, Andrew Stuart, Kaushik Bhattacharya, and Anima Anandkumar.
\newblock Multipole graph neural operator for parametric partial differential equations.
\newblock {\em Advances in Neural Information Processing Systems}, 33:6755--6766, 2020.

\bibitem[\protect\citeauthoryear{Li \bgroup \em et al.\egroup }{2021}]{li20}
Zongyi Li, Nikola~Borislavov Kovachki, Kamyar Azizzadenesheli, Kaushik Bhattacharya, Andrew Stuart, Anima Anandkumar, et~al.
\newblock Fourier neural operator for parametric partial differential equations.
\newblock In {\em International Conference on Learning Representations}, 2021.

\bibitem[\protect\citeauthoryear{Lillicrap \bgroup \em et al.\egroup }{2015}]{lillicrap2015continuous}
Timothy~P Lillicrap, Jonathan~J Hunt, Alexander Pritzel, Nicolas Heess, Tom Erez, Yuval Tassa, David Silver, and Daan Wierstra.
\newblock Continuous control with deep reinforcement learning.
\newblock {\em arXiv preprint arXiv:1509.02971}, 2015.

\bibitem[\protect\citeauthoryear{Lions and Magenes}{2012}]{lions12}
Jacques~Louis Lions and Enrico Magenes.
\newblock {\em Non-homogeneous boundary value problems and applications: Vol. 1}, volume 181.
\newblock Springer Science \& Business Media, 2012.

\bibitem[\protect\citeauthoryear{Lu \bgroup \em et al.\egroup }{2021}]{l21}
Lu~Lu, Pengzhan Jin, Guofei Pang, Zhongqiang Zhang, and George~Em Karniadakis.
\newblock Learning nonlinear operators via deeponet based on the universal approximation theorem of operators.
\newblock {\em Nature machine intelligence}, 3(3):218--229, 2021.

\bibitem[\protect\citeauthoryear{McCauley}{2013}]{mccauley2013stochastic}
Joseph~L McCauley.
\newblock {\em Stochastic calculus and differential equations for physics and finance}.
\newblock Cambridge University Press, 2013.

\bibitem[\protect\citeauthoryear{Mikulevicius and Rozovskii}{2004}]{m04}
Remigijus Mikulevicius and Boris~L Rozovskii.
\newblock Stochastic navier--stokes equations for turbulent flows.
\newblock {\em SIAM Journal on Mathematical Analysis}, 35(5):1250--1310, 2004.

\bibitem[\protect\citeauthoryear{O'shea and Nash}{2015}]{o2015introduction}
Keiron O'shea and Ryan Nash.
\newblock An introduction to convolutional neural networks.
\newblock {\em arXiv preprint arXiv:1511.08458}, 2015.

\bibitem[\protect\citeauthoryear{Pakravan \bgroup \em et al.\egroup }{2021}]{p21}
Samira Pakravan, Pouria {A. Mistani}, Miguel~A. Aragon-Calvo, and Frederic Gibou.
\newblock Solving inverse-pde problems with physics-aware neural networks.
\newblock {\em Journal of Computational Physics}, 440:110414, 2021.

\bibitem[\protect\citeauthoryear{Pirmorad \bgroup \em et al.\egroup }{2021}]{pirmorad2021deep}
Erfan Pirmorad, Faraz Khoshbakhtian, Farnam Mansouri, and Amir-massoud Farahmand.
\newblock Deep reinforcement learning for online control of stochastic partial differential equations.
\newblock {\em arXiv preprint arXiv:2110.11265}, 2021.

\bibitem[\protect\citeauthoryear{Raissi \bgroup \em et al.\egroup }{2019}]{r19}
Maziar Raissi, Paris Perdikaris, and George~E Karniadakis.
\newblock Physics-informed neural networks: A deep learning framework for solving forward and inverse problems involving nonlinear partial differential equations.
\newblock {\em Journal of Computational physics}, 378:686--707, 2019.

\bibitem[\protect\citeauthoryear{R{\"u}emelin}{1982}]{r82}
W~R{\"u}emelin.
\newblock Numerical treatment of stochastic differential equations.
\newblock {\em SIAM Journal on Numerical Analysis}, 19(3):604--613, 1982.

\bibitem[\protect\citeauthoryear{Salvi and Lemercier}{2021}]{c21}
Cristopher Salvi and Maud Lemercier.
\newblock Neural stochastic partial differential equations.
\newblock {\em CoRR}, abs/2110.10249, 2021.

\bibitem[\protect\citeauthoryear{Salvi \bgroup \em et al.\egroup }{2022}]{n22}
Cristopher Salvi, Maud Lemercier, and Andris Gerasimovics.
\newblock Neural stochastic pdes: Resolution-invariant learning of continuous spatiotemporal dynamics.
\newblock In S.~Koyejo, S.~Mohamed, A.~Agarwal, D.~Belgrave, K.~Cho, and A.~Oh, editors, {\em Advances in Neural Information Processing Systems}, volume~35, pages 1333--1344. Curran Associates, Inc., 2022.

\bibitem[\protect\citeauthoryear{Schulman \bgroup \em et al.\egroup }{2017}]{schulman2017proximal}
John Schulman, Filip Wolski, Prafulla Dhariwal, Alec Radford, and Oleg Klimov.
\newblock Proximal policy optimization algorithms.
\newblock {\em arXiv preprint arXiv:1707.06347}, 2017.

\bibitem[\protect\citeauthoryear{Sritharan}{1996}]{s96}
SS~Sritharan.
\newblock Nonlinear filtering of stochastic navier-stokes equation.
\newblock In {\em Nonlinear Stochastic PDEs: Hydrodynamic Limit and Burgers’ Turbulence}, pages 247--260. Springer, 1996.

\bibitem[\protect\citeauthoryear{Tleubergenov and Ibraeva}{2019}]{t19}
MI~Tleubergenov and GT~Ibraeva.
\newblock On the solvability of the main inverse problem for stochastic differential systems.
\newblock {\em Ukrainian Mathematical Journal}, 71(1):157--166, 2019.

\bibitem[\protect\citeauthoryear{Tripura and Chakraborty}{2022}]{t22}
Tapas Tripura and Souvik Chakraborty.
\newblock Wavelet neural operator: a neural operator for parametric partial differential equations.
\newblock {\em arXiv preprint arXiv:2205.02191}, 2022.

\bibitem[\protect\citeauthoryear{Wei \bgroup \em et al.\egroup }{2024}]{wei2024generative}
Long Wei, Peiyan Hu, Ruiqi Feng, Haodong Feng, Yixuan Du, Tao Zhang, Rui Wang, Yue Wang, Zhi-Ming Ma, and Tailin Wu.
\newblock Diffphycon: A generative approach to control complex physical systems.
\newblock In {\em The Thirty-eighth Annual Conference on Neural Information Processing Systems}, 2024.

\bibitem[\protect\citeauthoryear{Yang \bgroup \em et al.\egroup }{2021}]{y21}
Liu Yang, Xuhui Meng, and George~Em Karniadakis.
\newblock B-pinns: Bayesian physics-informed neural networks for forward and inverse pde problems with noisy data.
\newblock {\em Journal of Computational Physics}, 425:109913, 2021.

\bibitem[\protect\citeauthoryear{Yu and others}{2018}]{y18}
Bing Yu et~al.
\newblock The deep ritz method: a deep learning-based numerical algorithm for solving variational problems.
\newblock {\em Communications in Mathematics and Statistics}, 6(1):1--12, 2018.

\bibitem[\protect\citeauthoryear{Yu \bgroup \em et al.\egroup }{2022}]{y22}
Jeremy Yu, Lu~Lu, Xuhui Meng, and George~Em Karniadakis.
\newblock Gradient-enhanced physics-informed neural networks for forward and inverse pde problems.
\newblock {\em Computer Methods in Applied Mechanics and Engineering}, 393:114823, 2022.

\bibitem[\protect\citeauthoryear{Zhang \bgroup \em et al.\egroup }{2022}]{zhang2022neural}
Jingdong Zhang, Qunxi Zhu, and Wei Lin.
\newblock Neural stochastic control.
\newblock {\em Advances in neural information processing systems}, 35:9098--9110, 2022.

\bibitem[\protect\citeauthoryear{Zhang \bgroup \em et al.\egroup }{2023}]{zhang2023controlgym}
Xiangyuan Zhang, Weichao Mao, Saviz Mowlavi, Mouhacine Benosman, and Tamer Ba{\c{s}}ar.
\newblock Controlgym: Large-scale control environments for benchmarking reinforcement learning algorithms.
\newblock {\em arXiv preprint arXiv:2311.18736}, 2023.

\bibitem[\protect\citeauthoryear{Zhao \bgroup \em et al.\egroup }{2022}]{q22}
Qingqing Zhao, David~B. Lindell, and Gordon Wetzstein.
\newblock Learning to solve pde-constrained inverse problems with graph networks.
\newblock In {\em ICML}, 2022.

\bibitem[\protect\citeauthoryear{Zhou and Lu}{2024}]{zhou2024solving}
Mo~Zhou and Jianfeng Lu.
\newblock Solving time-continuous stochastic optimal control problems: Algorithm design and convergence analysis of actor-critic flow.
\newblock {\em arXiv preprint arXiv:2402.17208}, 2024.

\end{thebibliography}

\clearpage
\appendix

\section*{Technical Appendix}

\section{Model Architecture}
We provide the base model's architecture in Figure \ref{fig:model}. From it, we can see that the output $u_{t+1}$ is taken to calculate the transition loss, while the output $u_t$ and $f_t$ are used to calculate the reconstruction losses.

 As for CNN, all the blocks are combinations of MLPs and convolutional layers. And for FNO, all the blocks are combinations of MLPs and Fourier layers. Besides, we take the MLP as the policy net.

\begin{figure}[b]
  \centering
  \includegraphics[width=0.8\columnwidth]{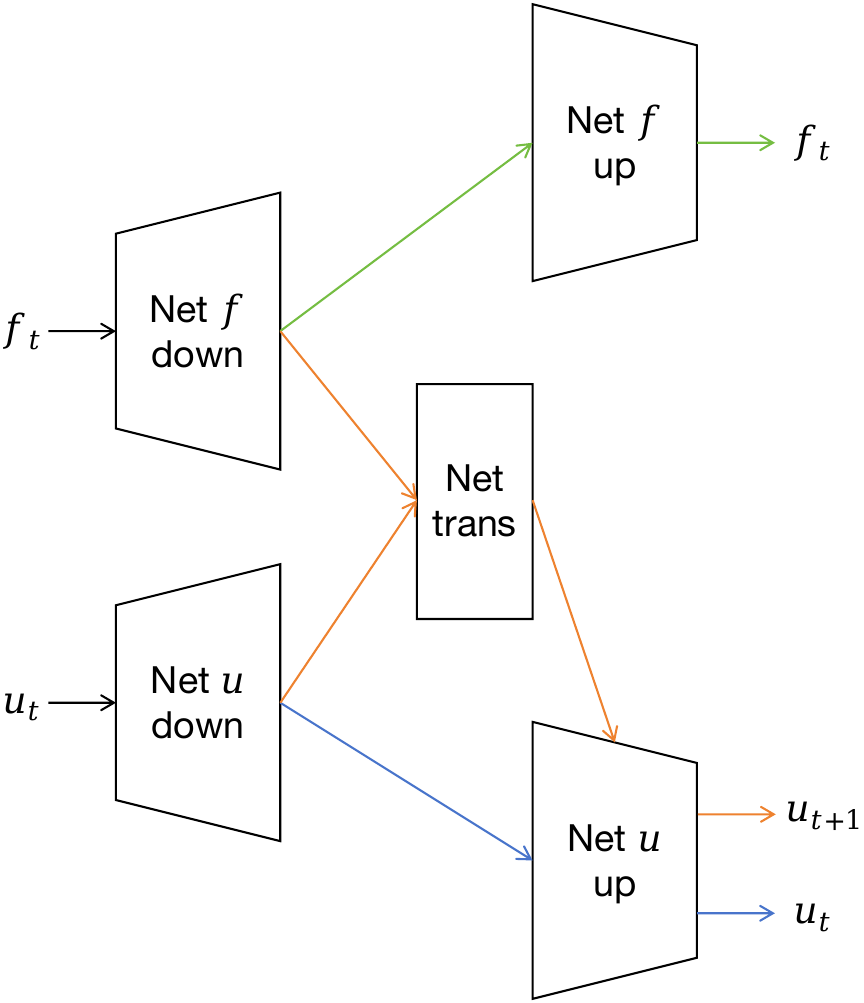}
  \caption{\textbf{Base Model Architecture.}}
  \label{fig:model}
\end{figure}

\section{Hyperparameters}
The detailed values of hyperparameters are reported in Table \ref{tab:1d_model_parameter}, Table \ref{tab:1d_policy_parameter}, Table \ref{tab:2d_model_parameter} and Table \ref{tab:2d_policy_parameter}. For each hyperparameter, we search for and finally use the best one.

\begin{table}[b]
  \begin{center}
    \begin{tabular}{l|l|l|l|l} %
    \multicolumn{5}{l}{}\\
    \midrule
      \text {Hyperparameter} & {CNN} & {RF-CNN} & {FNO} & {RF-FNO} \\
      \midrule
      Learning rate & 1e-3 &1e-3 &1e-2 &1e-2 \\
      Batch size & 64 &64 &25 &25 \\
      Number of epochs & 1000 &1000 &1000 &1000 \\
      \bottomrule
    \end{tabular}
  \end{center}
\caption{\textbf{Hyperparameters of 1D models' training.} }
\label{tab:1d_model_parameter}
\end{table}

\begin{table}
  \begin{center}
    \begin{tabular}{l|l|l|l|l} %
    \multicolumn{5}{l}{}\\
    \midrule
      \text {Hyperparameter} & {CNN} & {RF-CNN} & {FNO} & {RF-FNO} \\
      \midrule
      Learning rate & 2e-4 &2e-4 &5e-3 &5e-3 \\
      Batch size & 64 &64 &25 &25 \\
      Number of epochs & 1000 &1000 &1000 &1000 \\
      \bottomrule
    \end{tabular}
  \end{center}
\caption{\textbf{Hyperparameters of 2D models' training.} }
\label{tab:2d_model_parameter}
\end{table}

\begin{table}
  \begin{center}
    \begin{tabular}{l|l|l|l|l} %
    \multicolumn{5}{l}{}\\
    \midrule
      \text {Hyperparameter} & {CNN} & {RF-CNN} & {FNO} & {RF-FNO} \\
      \midrule
       \multicolumn{1}{l|}{Hidden nodes} & \multicolumn{4}{l}{257, 2048, 1024, 1024, 64}\\
      \midrule
       Learning rate&5e-5 &5e-4 &1e-5 &5e-5 \\
       Batch size&1 &100 &400 &400 \\
       Number of epochs&3 &128 &80 &80 \\
      \bottomrule
    \end{tabular}
  \end{center}
\caption{\textbf{Hyperparameters of 1D policy net's training.} }
\label{tab:1d_policy_parameter}
\end{table}

\begin{table}
  \begin{center}
    \begin{tabular}{l|l|l|l|l} %
    \multicolumn{5}{l}{}\\
    \midrule
      \text {Hyperparameter} & {CNN} & {RF-CNN} & {FNO} & {RF-FNO} \\
      \midrule
       Learning rate&8e-5 &8e-5 &4e-4 &5e-4 \\
       Batch size&32 &32 &11 &11 \\
       Number of epochs&15 &160 &10 &6 \\
      \bottomrule
    \end{tabular}
  \end{center}
\caption{\textbf{Hyperparameters of 2D policy net's training.} }
\label{tab:2d_policy_parameter}
\end{table}

\section{Results Visualizations}

In Figure \ref{fig:rd_full}, we present additional visualized results from the 1D reaction- diffusion experiments. It can be observed that MB-CC consistently achieves significantly better performance across all tasks in the figure.

\begin{figure*}
  \centering
  \makebox[\textwidth][c]{\includegraphics[width=2\columnwidth]{icml2024/fig/rd.pdf}}
  \makebox[\textwidth][c]{\includegraphics[width=2\columnwidth]{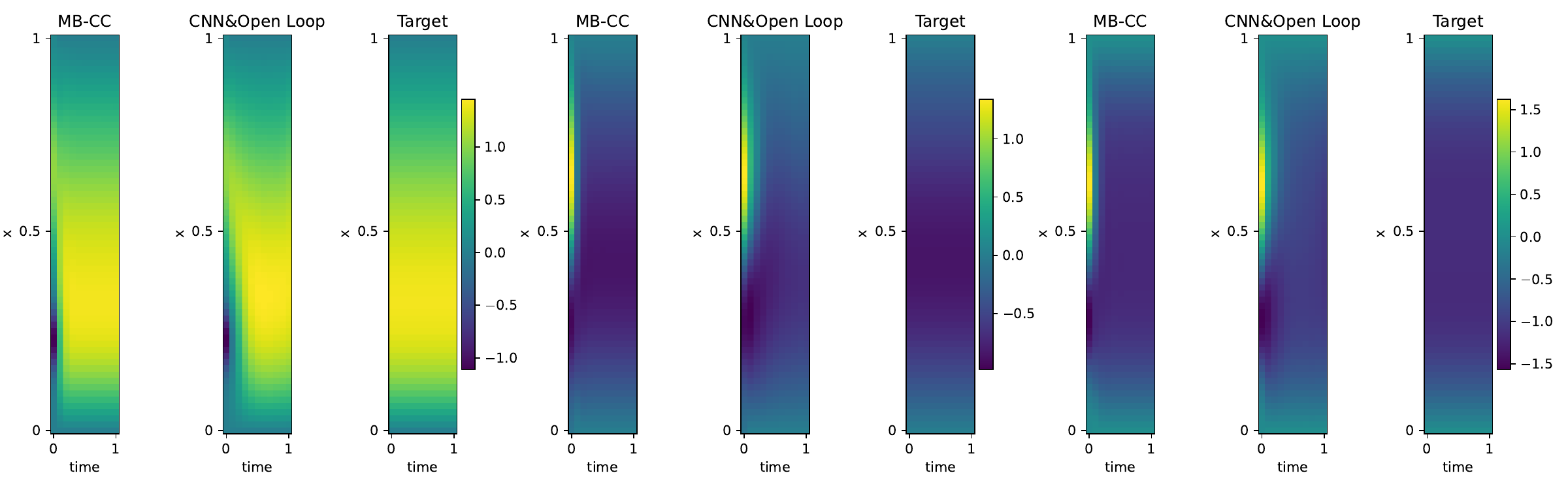}} 
  \makebox[\textwidth][c]{\includegraphics[width=2\columnwidth]{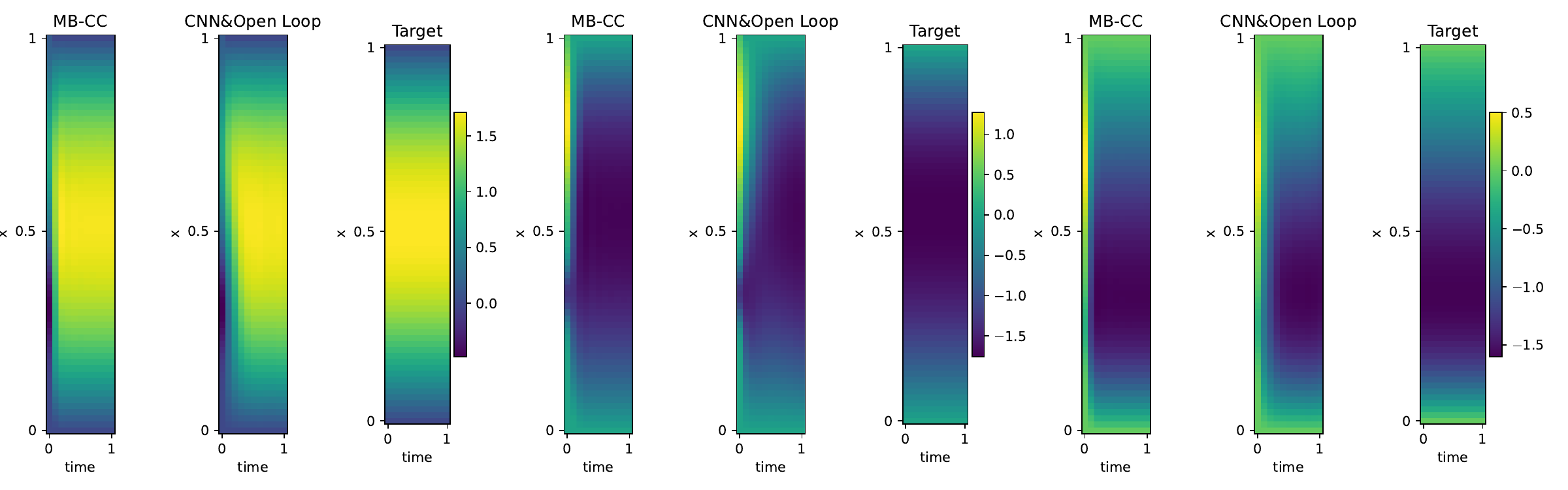}}  
  \makebox[\textwidth][c]{\includegraphics[width=2\columnwidth]{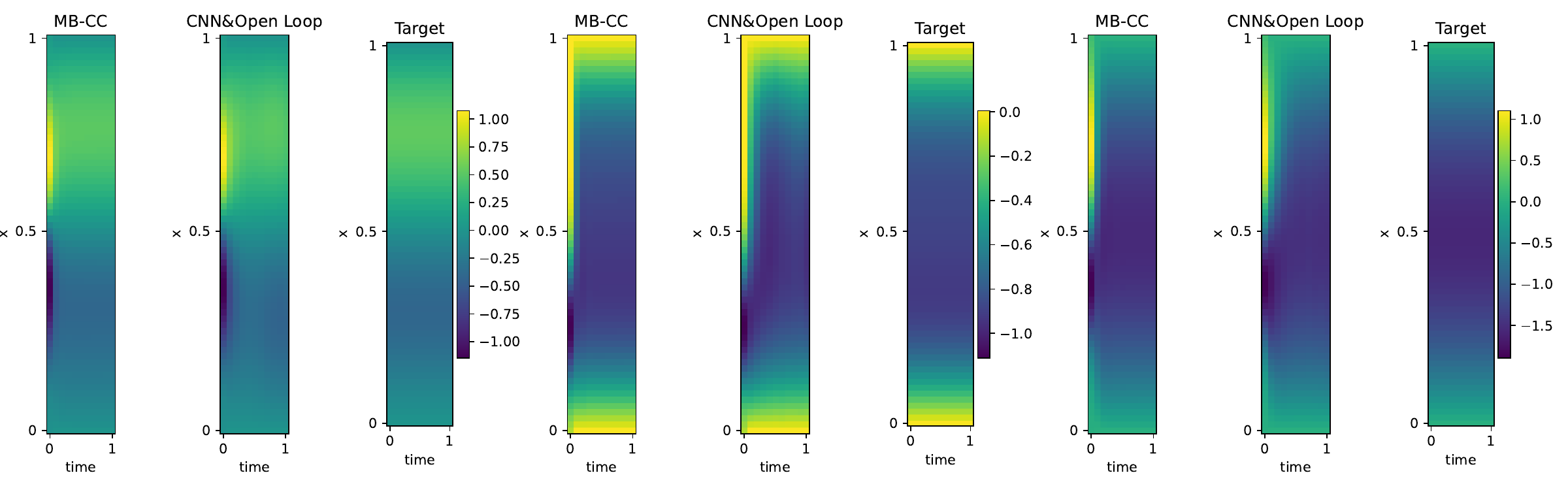}}
  \caption{\textbf{Visualizations of 1D experiments.}}
  \label{fig:rd_full}
\end{figure*}

\section{Discretized Operator $\mathcal{L}^{\text{dis}}$}
\label{app:l_dis}
The discretization of the operator we use is based on the first-order finite difference method. For example, in the 1-D stochastic reaction- diffusion equation, $\mathcal{L}$ equals the 1d Laplace operator, i.e., $\mathcal{L}=\Delta$, the discretized operator $\mathcal{L}^{\text{dis}}$ we take is 
\begin{align}
\Delta^{\text{dis}}=\frac{1}{\epsilon^2}\left(\begin{matrix}-2 &1 &0&\cdots&0\\
1&-2&1&\ddots&0\\
\vdots&\ddots&\ddots&\ddots&\vdots\\
0&\ddots&1&-2&1\\
0&\cdots&0&1&-2\end{matrix}\right),
\end{align}where $\epsilon$ is the discretized step. 

\section{Computational Resources}

All the experiments are conducted on a single Tesla-A100 GPU with 80GB memory. The training time of different models is quite different, while the introduction of the RF block almost maintains the efficiency. The detailed inference time can be found in the Experiment section.

\end{document}